\input harvmac
\noblackbox

\newcount\figno
\figno=0
\def\fig#1#2#3{
\par\begingroup\parindent=0pt\leftskip=1cm\rightskip=1cm\parindent=0pt
\baselineskip=11pt

\global\advance\figno by 1
\midinsert
\epsfxsize=#3
\centerline{\epsfbox{#2}}
\vskip 12pt
\centerline{{\bf Figure \the\figno:} #1}\par
\endinsert\endgroup\par}
\def\figlabel#1{\xdef#1{\the\figno}}

\def\risunok#1#2#3{\vskip 15pt
\global\advance\figno by 1
\centerline{\epsfbox{#1}}
\vskip 10pt
\centerline{{\bf Fig. #3: } #2}
\vskip 15pt }
\def\kartinka#1#2#3{
\par\begingroup\parindent=0pt\leftskip=1cm\rightskip=1cm\parindent=0pt
\baselineskip=11pt

\global\advance\figno by 1
\midinsert
\centerline{\epsfbox{#1}}
\vskip 12pt
\centerline{{\bf Fig. #3: }#2}\par
\endinsert\endgroup\par}

\def\np#1#2#3{Nucl. Phys. {\bf B#1} (#2) #3}
\def\pl#1#2#3{Phys. Lett. {\bf B#1} (#2) #3}
\def\prl#1#2#3{Phys. Rev. Lett.{\bf #1} (#2) #3}
\def\physrev#1#2#3{Phys. Rev. {\bf D#1} (#2) #3}


\font\cmss=cmss10
\font\cmsss=cmss10 at 7pt
\def\rlx{\relax\leavevmode}
\def\inbar{\vrule height1.5ex width.4pt depth0pt}
\def\IC{\relax\,\hbox{$\inbar\kern-.3em{\rm C}$}}
\def\IN{\relax{\rm I\kern-.18em N}}
\def\IP{\relax{\rm I\kern-.18em P}}
\def\ZZ{\rlx\leavevmode\ifmmode\mathchoice{\hbox{\cmss Z\kern-.4em Z}}
  {\hbox{\cmss Z\kern-.4em Z}}{\lower.9pt\hbox{\cmsss Z\kern-.36em Z}}
  {\lower1.2pt\hbox{\cmsss Z\kern-.36em Z}}\else{\cmss Z\kern-.4em
  Z}\fi}
\def\IZ{\relax\ifmmode\mathchoice
{\hbox{\cmss Z\kern-.4em Z}}{\hbox{\cmss Z\kern-.4em Z}}
{\lower.9pt\hbox{\cmsss Z\kern-.4em Z}}
{\lower1.2pt\hbox{\cmsss Z\kern-.4em Z}}\else{\cmss Z\kern-.4em
Z}\fi}

\def\narrowplus{\kern -.04truein + \kern -.03truein}
\def\narrowminus{- \kern -.04truein}
\def\narrowminussub{\kern -.02truein - \kern -.01truein}

\def\bps{Bogomol'nyi}

\def\kh{K\"{a}hler }
\def\half{{1\over 2}}

\def\g{{\gamma}}

\def\r{{\rightarrow}}

\def\frac#1#2{{#1\over #2}}

\def\CC{{\cal C}}

\def\CP{{\cal P}}

\def\IZ{\relax\ifmmode\mathchoice
{\hbox{\cmss Z\kern-.4em Z}}{\hbox{\cmss Z\kern-.4em Z}}
{\lower.9pt\hbox{\cmsss Z\kern-.4em Z}}
{\lower1.2pt\hbox{\cmsss Z\kern-.4em Z}}\else{\cmss Z\kern-.4em
Z}\fi}
\def\IB{\relax{\rm I\kern-.18em B}}
\def\IC{{\relax\hbox{$\inbar\kern-.3em{\rm C}$}}}
\def\ID{\relax{\rm I\kern-.18em D}}
\def\IE{\relax{\rm I\kern-.18em E}}
\def\IF{\relax{\rm I\kern-.18em F}}
\def\IG{\relax\hbox{$\inbar\kern-.3em{\rm G}$}}
\def\IGa{\relax\hbox{${\rm I}\kern-.18em\Gamma$}}
\def\IH{\relax{\rm I\kern-.18em H}}
\def\II{\relax{\rm I\kern-.18em I}}
\def\IK{\relax{\rm I\kern-.18em K}}
\def\IP{\relax{\rm I\kern-.18em P}}

\def\p{\partial}

\font\cmss=cmss10 \font\cmsss=cmss10 at 7pt
\def\IR{\relax{\rm I\kern-.18em R}}

\def\S{{\Sigma}}

%

%
%
\def\eqnn#1{\xdef #1{(\secsym\the\meqno)}\writedef{#1\leftbracket#1}%
\global\advance\meqno by1\wrlabeL#1}
\def\eqna#1{\xdef #1##1{\hbox{$(\secsym\the\meqno##1)$}}
\writedef{#1\numbersign1\leftbracket#1{\numbersign1}}%
\global\advance\meqno by1\wrlabeL{#1$\{\}$}}
\def\eqn#1#2{\xdef #1{(\secsym\the\meqno)}\writedef{#1\leftbracket#1}%
\global\advance\meqno by1$$#2\eqno#1\eqlabeL#1$$}


\input epsf

\def\SUSY#1{{{\cal N}= {#1}}}                   




\def\CP#1{{{\bf P}^{#1}}}               












\mathchardef\subset="321A




\def\Arg{{\rm Arg  }}



\lref\rschw{J. Schwartz, hep-th/9607201\semi
O. Aharony, J. Sonnenschein and S. Yankielowicz, \np{474}{1996}{309},
hep-th/9603009\semi
M. Gaberdiel and B. Zwiebach. hep-th/9709013.}

\lref\rqthree{K. Dasgupta and S. Mukhi, hep-th/9711094.}

\lref\rmthree{M. Krogh and S. Lee, hep-th/9712050\semi Y. Matsuo and K.
Okuyama, hep-th/9712070.}

\lref\rmseven{I. Kishimoto and N. Sasakura, hep-th/9712180.}

\lref\rima{Y. Imamura, hep-th/9802189.}

\lref\rFtheory{C. Vafa, hep-th/9602022, \np{469}{1996} 403.}

\lref\rbilal{A. Bilal and F. Ferrari,  hep-th/9602082, \np{469}{1996} 387;
 hep-th/9605101, \np{480}{1996} 589; hep-th/9606192, Nucl. Phys. Proc.
Suppl. {\bf 52A} 305, 1997; hep-th/9706145. }

\lref\rsen{A. Sen, hep-th/9605150, \np{475}{1996} 562; hep-th/9608005,
 Phys. Rev. {\bf D55} (1997) 2501.}

\lref\rprobe{T. Banks, M. Douglas and N. Seiberg,  hep-th/9605199,
\pl{387}{1996} 287.}

\lref\randrei{A. Mikhailov, hep-th/9708068.}

\lref\rpiljin{M. Henningson and P. Yi, hep-th/9707251, \physrev{57}{1998}
1291.}

\lref\rggm{A.~Gorsky, S.~Gukov and A.~Mironov, hep-th/9707120.}

\lref\rsd{A. Klemm, W. Lerche, P. Mayr, C. Vafa and N. Warner, hep-th/9604034,
\np{477}{1996} 746.}

\lref\rmoresd{A. Brandhuber and S. Stieberger, hep-th/9610053, \np{488}{1997}
199\semi
J. Schulze and N. Warner, hep-th/9702012\semi J. Rabin, hep-th/9703145.}

\lref\rwittenM{E. Witten, hep-th/9703166, \np{500}{1997} 3.}

\lref\rmukhi{K. Dasgupta and S. Mukhi, hep-th/9711094.}

\lref\rCV{S. Cecotti and C. Vafa, hep-th/9211097, Comm. math. Phys. {\bf 158}
(1993) 596.}

\lref\rSW{N. Seiberg and E. Witten, hep-th/9407087, \np{429}{1994} 19;
hep-th/9408099,
\np{431}{1994} 484.}

\lref\rthesis{S. Sethi, Ph. D. Thesis, Harvard University (1996) chapter 3.}

\lref\rmans{M. Henningson, hep-th/9510138, \np{461}{1996} 101.}

\lref\rgeom{S. Katz, P. Mayr and C. Vafa, hep-th/9706110.}

\lref\rsunil{K. Dasgupta and S. Mukhi, hep-th/9606044, \pl{385}{1996}{125}. }

\lref\roren{O. Bergman and A. Fayyazuddin, hep-th/9802033. }

\lref\rfourfold{S. Sethi, C. Vafa and E. Witten, hep-th/9606122,
\np{480}{1996}{213}.}

\lref\rmonopole{S. Sethi, M. Stern and E. Zaslow, hep-th/9508117,
\np{457}{1995}{484}\semi
J. Gauntlett and J. Harvey, hep-th/9508156, \np{463}{1996}{287}.}

\lref\rfay{A. Fayyazuddin, hep-th/9701185, \np{497}{1997}{101}.}

\lref\rzwie{M. Gaberdiel, T. Hauer and B. Zwiebach, hep-th/9801205.}

\lref\rmIIB{P. Aspinwall, ``Some Relationships Between Dualities in
String Theory,'' hep-th/9508154, Nucl. Phys. Proc. Suppl. {\bf 46}
(1996) 30\semi J. Schwarz, ``The Power of M Theory,''
hep-th/9510086, \pl{367}{1996}{97}. }

\lref\rfay{A. Fayyazuddin and M. Spalinski, hep-th/9706087.}

\lref\rnatiprobe{N. Seiberg, hep-th/9606017, \pl{384}{1996}{81}.}

\lref\rnak{T. Nakatsu, K. Ohta, T. Yokono and Y. Yoshida, hep-th/9711117\semi
Y. Yoshida,
hep-th/9711177.}

\lref\rgreen{C. Bachas and M. Green, hep-th/9712187.}

\lref\rorenym{O. Bergman, hep-th/9712211.}

\lref\rymsolution{K. Hashimoto, H. Hata and N. Sasakura, hep-th/9803127.}

\lref\rfivebranes{B. Kol and J. Rahmfeld, hep-th/9801067\semi
O. Aharony, A. Hanany and B. Kol, hep-th/9710116.}

\lref\ransar{ A. Fayyazuddin, hep-th/9701185, \np{497}{1997}{101}.}

\lref\rsoojong{S.-J. Rey and J.-T. Lee, hep-th/9711202.}

\lref\rcallan{C. Callan, Jr. and L. Thorlacius, hep-th/9803097.}

\lref\rmike{M. Douglas, D. Lowe and J. Schwarz, hep-th/9612062, 
\pl{394}{1997}{297}. }

\lref\rpyi{K. Lee and P. Yi, hep-th/9706023.}

\lref\rand{A. Johansen, hep-th/9608186, \pl{395}{1997}{36}.}

\lref\rnik{A. Lawrence and N. Nekrasov, hep-th/9706025.}

\lref\rpol{J. Polchinski, ``TASI Lectures on D-Branes,''
    hep-th/9611050\semi J. Polchinski, S. Chaudhuri and C. Johnson,
    ``Notes on D-Branes,'' hep-th/9602052. }

\lref\rBFSS{T. Banks, W. Fischler, S. H. Shenker, and L. Susskind, ``M
    Theory As A Matrix Model: A Conjecture,''
    hep-th/9610043, Phys. Rev. {\bf D55} (1997) 5112.}

\lref\rwtensor{E. Witten, ``Some Comments on String Dynamics,''
    hep-th/9507121.}

\lref\rstensor{A. Strominger, ``Open P-Branes,'' hep-th/9512059,
    \pl{383}{1996}{44}.}

\lref\rsdecoupled{N. Seiberg, ``New Theories in Six Dimensions and
    Matrix Description of M-theory on $T^5$ and $T^5/\IZ_2$,''
    hep-th/9705221.}

\lref\rashoke{ A. Sen, ``A Note on Enhanced Gauge Symmetries in
    M and String Theory,'' hep-th/9707123; ``Dynamics of Multiple
    Kaluza-Klein Monopoles in M and String Theory,'' hep-th/9707042.}

\lref\kutetal{D. Berenstein, R. Corrado and J. Distler, ``On the
    Moduli Spaces of M(atrix)-Theory Compactifications,''
    hep-th/9704087\semi  S. Elitzur, A. Giveon, D. Kutasov and
    E. Rabinovici, ``Algebraic Aspects of Matrix Theory on $T^d$,''
    hep-th/9707217.}

\lref\rtwoform{J. P. Gauntlett and D. Lowe, ``Dyons and S-Duality in
    N=4 Supersymmetric Gauge Theory,'' hep-th/9601085,
    \np{472}{1996}{194}\semi K. Lee, E. Weinberg and P. Yi,
    ``Electromagnetic Duality and $SU(3)$ Monopoles,'' hep-th/9601097,
    \pl{376}{1996}{97}.}

\lref\rmoore{A. Losev, G. Moore, and S. Shatashvili, ``M \& m's ,''
    hep-th/9707250.}

\lref\rbrunner{I. Brunner and A. Karch, ``Matrix Description of
    M-theory on $T^6$,'' hep-th/9707259.}

\lref\rSeiWHY{N. Seiberg, ``Why is the Matrix Model Correct?,''
hep-th/9710009.}

\lref\rSenTn{A. Sen,
  ``D0 Branes on $T^n$ and Matrix Theory,'' hep-th/9709220.}

\lref\rDVV{R. Dijkgraaf, E. Verlinde and H. Verlinde, ``BPS Spectrum
    of the Five-Brane and Black Hole Entropy,'' hep-th/9603126,
    \np{486}{1997}{77}; ``BPS Quantization of the Five-Brane,''
    hep-th/9604055, \np{486}{1997}{89}.}

\lref\rextraDVV{R. Dijkgraaf, E. Verlinde and H. Verlinde, ``5D Black Holes and
Matrix Strings,'' hep-th/9704018.}

\lref\rsixbrane{P. Townsend, ``The Eleven Dimensional Supermembrane
    Revisited,'' hep-th/9501068, \pl{350}{1995}{184}.}

\lref\rmultitn{R. Sorkin, ``Kaluza-Klein Monopole,''
    \prl{51}{1983}{87}\semi D. Gross and M. Perry, ``Magnetic Monopoles
    in Kaluza-Klein Theories,'' \np{226}{1983}{29}.}

\lref\rmIIB{P. Aspinwall, ``Some Relationships Between Dualities in
    String Theory,'' hep-th/9508154, Nucl. Phys. Proc. Suppl. {\bf 46}
    (1996) 30\semi J. Schwarz, ``The Power of M Theory,''
    hep-th/9510086, \pl{367}{1996}{97}. }

\lref\rgeneralsixbrane{ C. Hull,  ``Gravitational Duality, Branes and
    Charges,'' hep-th/9705162\semi E. Bergshoeff, B. Janssen, and
    T. Ortin, ``Kaluza-Klein Monopoles and Gauged Sigma Models,''
    hep-th/9706117\semi Y. Imamura, ``Born-Infeld Action and
Chern-Simons
    Term {}from Kaluza-Klein Monopole in M-theory,'' hep-th/9706144.}

\lref\rbd{M. Berkooz and M. Douglas, ``Five-branes in M(atrix)
    Theory,'' hep-th/9610236, \pl{395}{1997}{196}.}

\lref\rbraneswith{M. Douglas, ``Branes within Branes,''
    hep-th/9512077.}

\lref\rquantumfive{O. Aharony, M. Berkooz, S. Kachru, N. Seiberg, and
    E. Silverstein, ``Matrix Description of Interacting Theories in Six
    Dimensions,'' hep-th/9707079.}

\lref\rstringfive{E. Witten, ``On The Conformal Field Theory of The
    Higgs Branch,'' hep-th/9707093.}
    \lref\rSS{S. Sethi and L. Susskind, ``Rotational Invariance in the
    M(atrix) Formulation of Type IIB Theory,'' hep-th/9702101,
    \pl{400}{1997}{265}.}

\lref\rBS{T. Banks and N. Seiberg, ``Strings from Matrices,''
    hep-th/9702187, \np{497}{1997}{41}.}

\lref\rgilad{A. Hanany and G. Lifschytz, ``M(atrix) Theory on $T^6$
    and a m(atrix) Theory Description of KK Monopoles,''
    hep-th/9708037.}

\lref\rreview{N. Seiberg, ``Notes on Theories with 16
    Supercharges,'' hep-th/9705117.}

\lref\rDVVstring{R. Dijkgraaf, E. Verlinde and H. Verlinde, ``Matrix
    String Theory,'' hep-th/9703030.}
    \lref\rprobes{M. Douglas, ``Gauge Fields and D-branes,''
    hep-th/9604198.}

\lref\rdoumoo{M. Douglas and G. Moore, ``D-Branes,
    Quivers, and ALE Instantons,'' $\,$ hep-th/9603167\semi
    C. Johnson and R. Myers, ``Aspects of Type IIB Theory on ALE
    Spaces,'' hep-th/9610140, \physrev{55}{1997}{6382}.}

\lref\rdouglas{M. Douglas, ``Enhanced Gauge Symmetry in M(atrix)
    Theory,'' hep-th/9612126\semi W. Fischler and A. Rajaraman,
 ``M(atrix) String Theory on K3,'' hep-th/9704123.}

\lref\rsprobes{N. Seiberg, ``Gauge Dynamics And Compactification To
    Three Dimensions,'' hep-th/9607163, \pl{384}{1996}{81}}

\lref\rswthree{N. Seiberg and E. Witten, ``Gauge Dynamics and
    Compactifications to Three Dimensions,'' hep-th/9607163.}

\lref\rthroat{D.-E. Diaconescu and N. Seiberg, ``The Coulomb Branch of
    $(4,4)$ Supersymmetric Field Theories in Two Dimensions,''
    hep-th/9707158. }

\lref\rTduality{T. Banks, M. Dine, H. Dykstra and W. Fischler,
    ``Magnetic Monopole Solutions of String Theory,''
    \pl{212}{1988}{45}\semi C. Hull and P. Townsend, ``Unity of
    Superstring Dualities,'' hep-th/9410167, \np{438}{109}{1995}\semi
    H. Ooguri, C. Vafa, ``Two Dimensional Black Hole and Singularities
    of Calabi-Yau Manifolds,'' Nucl.Phys. {\bf B463} (1996) 55,
    hep-th/9511164\semi D. Kutasov, ``Orbifolds and Solitons,''
    Phys.  Lett {\bf B383} (1996) 48, hep-th/9512145\semi
    H. Ooguri and C. Vafa,
    ``Geometry of N=1 Dualities in Four Dimensions,'' hep-th/9702180.}

\lref\raps{P. Argyres, R. Plesser and N. Seiberg, ``The Moduli Space
    of N=2 SUSY QCD and Duality in N=1 SUSY QCD,'' hep-th/9603042,
    \np{471}{1996}{159}.}

\lref\rgms{O. Ganor, D. Morrison and N. Seiberg, ``Branes, Calabi-Yau
    Spaces, and Toroidal Compactification of the N=1 Six Dimensional
    $E_8$ Theory,'' hep-th/9610251, \np{487}{1997}{93}.}

\lref\rchs{ C.G. Callan, J.A. Harvey, A. Strominger, ``Supersymmetric
    String Solitons,'' hep-th/9112030, \np{359}{1991}{611}\semi
    S.-J. Rey, in ``The  Proc. of the Tuscaloosa Workshop
    1989,'' 291; Phys. Rev. {\bf D43} (1991) 526; S.-J. Rey, In DPF
    Conf. 1991, 876.}

\lref\rmotl{L. Motl, ``Proposals on nonperturbative superstring
    interactions,'' hep-th/9701025.}

\lref\rwati{W. Taylor IV, ``D-Brane Field Theory on Compact Space,''
    hep-th/9611042, \pl{394}{1997}{283}.}

\lref\rkin{K. Intriligator, ``New String Theories in Six Dimensions
    via Branes at Orbifold Singularities,'' hep-th/9708117.}

\lref\rmont{C. Montonen and D. Olive, \pl{72}{1977}{117}.}

\lref\rsdual{A. Sen, ``Dyon-Monopole Bound States, Self-Dual Harmonic
    Forms on the Multi-Monopole Moduli Space, and SL(2,Z) Invariance in
    String Theory,'' hep-th/9402032, \pl{329}{1994}{217}.}

\lref\rstrong{C. Vafa and E. Witten, ``A Strong Coupling Test of
    S-Duality,'' hep-th/9408074, \np{432}{1994}{3}. }

\lref\rlenny{L. Susskind, ``T Duality in M(atrix) Theory and S Duality
    in Field Theory,'' hep-th/9611164. }

\lref\rori{O. Ganor, S. Ramgoolam and W. Taylor IV, ``Branes, Fluxes and
    Duality in M(atrix)-Theory,'' hep-th/9611202, \np{492}{1997}{191}.}

\lref\rfourtorus{M. Rozali, ``Matrix Theory and U-Duality in Seven
    Dimensions,'' hep-th/9702136, Phys. Lett. {\bf B400} (1997) 260\semi
M, Berkooz, M. Rozali and N. Seiberg, ``Matrix Description of M-theory
on $T^4$ and $T^5$,'' hep-th/9704089.}

\lref\rnati{N. Seiberg and S. Sethi, ``Comments on Neveu-Schwarz
    Five-Branes,'' hep-th/9708085.}

\lref\rdlcq{L. Susskind, ``Another Conjecture about M(atrix) Theory,''
    hep-th/9704080.}

\lref\rsIIB{S. Sethi, ``The Matrix Formulation of Type IIB
    Five-Branes,'' hep-th/9710005.}

\lref\reIIB{E. Witten, ``New `Gauge' Theories in Six-Dimensions,''
    hep-th/9710065.}

\lref\rhmon{S. Sethi and M. Stern, ``A Comment on the Spectrum of
    H-Monopoles,'' hep-th/9607145, \pl{398}{1997}{47}.}

\lref\rconjecture{J. de Boer, K. Hori, H. Ooguri and Y. Oz, ``Mirror
    Symmetry in Three-Dimensional Gauge Theories, Quivers and D-branes,''
    hep-th/9611063, \np{493}{1997}{101}.}

\lref\rmirror{K. Intriligator and N. Seiberg, ``Mirror Symmetry in Three
    Dimensional Gauge Theories,'' hep-th/9607207, \pl{386}{1996}{513}. }

\lref\rBR{J. Brodie and S. Ramgoolam, ``On Matrix Models of M5 branes,''
hep-th/9711001.}

\lref\rjulie{J. D. Blum and K. Intriligator, ``Consistency Conditions for
Branes at
Orbifold Singularities,'' hep-th/9705030; ``New Phases of String Theory and 6d
RG
Fixed Points via Branes at Orbifold Singularities,'' hep-th/9705044.}

\lref\rImamura{Y. Imamura,
   ``A Comment on Fundamental Strings in M(atrix) Theory,''
  hep-th/9703077.}

\lref\rEGKRalg{S. Elitzur, A. Giveon, D. Kutasov, E. Rabinovici,
   ``Algebraic Aspects of Matrix Theory on $T^d$,''
  hep-th/9707217.}

\lref\rGivKut{A. Giveon and D. Kutasov, to appear.}

\lref\rHacVer{F. Hacquebord and H. Verlinde,
   ``Duality symmetry of $\SUSY{4}$ Yang-Mills theory on $T^3$,''
  hep-th/9707179.}

\lref\rWitBR{E. Witten,
  ``Solutions Of Four-Dimensional Field Theories Via M Theory,''
  hep-th/9703166.}

\lref\rtHooft{G. 't Hooft,
  ``On the phase transition towards permanent quark confinement,''
  \np{138}{1978}{1-25}.}

\lref\rHelPol{S. Hellerman and J. Polchinski,
  ``Compactification in the Light-Like Direction,''
  hep-th/9711037.}

\lref\rVafIOD{A. Sen, ``U-duality and Intersecting D-branes,''
hep-th/9511026\semi
 C. Vafa, ``Gas of D-Branes and Hagedorn Density of BPS States,''
hep-th/9511088.}

\lref\rGriHar{P. Griffiths and J. Harris,
  {\it Principles of Algebraic Geometry}, Wiley-Interscience, New York,
  1978.}

\lref\rGukov{S. Gukov,
  ``Seiberg-Witten Solution from Matrix Theory,'' hep-th/9709138.}

\lref\rFMW{R. Friedman, J. Morgan, and E. Witten,
   ``Vector Bundles And F Theory,'' hep-th/9701162\semi
M. Bershadsky, A. Johansen, T. Pantev and V. Sadov, ``On Four-Dimensional
Compactifications of F-Theory, '' hep-th/9701165.}

\lref\rEGKRalg{S. Elitzur, A. Giveon, D. Kutasov, E. Rabinovici,
  ``Algebraic Aspects of Matrix Theory on ${\bf T}^d$,''
  hep-th/9707217.}

\lref\rneqone{S. Elitzur, A. Giveon, D. Kutasov,
  {``Brane and $\SUSY{1}$ Duality In String Theory,''}
  hep-th/9702014.}


\lref\rmukai{S. Mukai, ``Duality between $D(X)$ and $D(
  \hat{X}$ ) with its Application to Picard Sheaves,'' Nagoya.
  Math. J. {\bf 81} (1981) 153.}

\lref\rKMV{S. Katz, P. Mayr and C. Vafa, ``Mirror Symmetry and Exact Solution
of 4D
  N=2 Gauge Theories - I,'' hep-th/9706110.}

\lref\rBarak{B. Kol, ``On 6d ``Gauge'' Theories with Irrational Theta Angle, ''
hep-th/
9711017.}

\lref\rDOS{M. Douglas, H. Ooguri and S. Shenker, ``Issues in M(atrix)
 Theory Compactification,'' hep-th/9702203.}

\lref\rBerRoz{S. Govindarajan, ``A note on M(atrix) theory in seven
dimensions with eight supercharges, '' hep-th/9705113\semi
M. Berkooz and M. Rozali, ``String dualities from M(atrix) Theory,''
hep-th/9705175.}

\lref\rSeiFDS{N. Seiberg,
  { ``Five Dimensional SUSY Field Theories, Non-trivial Fixed Points
  And String Dynamics,''} hep-th/9608111.}

\lref\rmayr{A. Klemm, W. Lerche, P. Mayr and C. Vafa, ``Self-Dual
Strings and N=2 Supersymmetric Field Theory,'' hep-th/9604034.}

\lref\ryi{K. Lee and P. Yi, ``Monopoles and Instantons on Partially
Compactified D-Branes,'' hep-th/9702107.}

\Title{\vbox{\hbox{hep--th/9803142}
                 \hbox{HUTP-98/A011}
\hbox{IASSNS--HEP--98/11}
\hbox{PUHEP-1765}
\hbox{NSF-ITP-98-028}
\hbox{ITEP-TH-11/98}}}
{Geometric Realizations of BPS States in N=2 Theories}
\smallskip
\centerline{Andrei Mikhailov\footnote{$^\ast$} {andrei@puhep1.princeton.edu},
Nikita Nekrasov\footnote{$^\spadesuit$}{nikita@string.harvard.edu} and Savdeep
Sethi\footnote{$^\dagger$} {sethi@sns.ias.edu}  }
\medskip\centerline{ $\ast$ \it Department of Physics, Princeton University,
 Princeton, NJ
08544, USA}
\centerline{$\spadesuit$ \it Lyman Laboratory, Harvard University, Cambridge,
 MA
02138, USA}

\centerline{ \it \phantom{$\spadesuit$} Institute for Theoretical and 
Experimental Physics, Moscow,  117259, Russia} 

\medskip\centerline{$\dagger$ \it School of Natural Sciences, Institute for
Advanced Study, Princeton, NJ 08540, USA}
\vskip .75in

We study the BPS spectrum of the theory on a D3-brane probe in F theory. The BPS states are realized by multi-string configurations in spacetime. Only certain configurations obeying a selection rule give rise to BPS states in the four-dimensional probe theory. Using these string configurations, we determine the spectrum of N=2 $SU(2)$ Yang-Mills. We also explore the relation between multi-string configurations, M theory membranes and self-dual strings.

\Date{3/98}

%
%

\newsec{Introduction}

One of the most striking phenomena found in certain supersymmetric theories is
the
non-analytic behavior of the BPS spectrum as the moduli of the theory are
varied. There
are generally hypersurfaces of real codimension one in the moduli space where
the BPS bound no longer
forbids the decay of BPS particles because some
of the complex central charges have aligned. When one of these curves of
marginal stability (CMS)
is crossed, some BPS particles may decay
into many particle states, or certain many particle states may bind to give a
new single particle
BPS state. This phenomena was originally observed in two-dimensional N=2
theories \rCV, and was
seen to be necessary in four-dimensional N=2 gauge theories \rSW.  The same
non-analytic behavior
will appear generally in string backgrounds preserving enough supersymmetry.

We will consider the case of N=2 four-dimensional theories. In some of these
theories, for example
$SU(2)$ Yang-Mills with a massive hypermultiplet,  certain decay processes can
be studied using
semi-classical techniques \refs{\rthesis, \rmans}. However, most curves of
marginal stability do
not extend to the region of weak coupling, so a direct analysis in Yang-Mills
is difficult. However,
in these cases, consistency of the proposed vacuum solution can often be used
to determine the BPS
spectrum inside and outside a curve of marginal stability \rbilal. This
technique can be extended
to theories whose curves are known, but which have no known Lagrangian
description like the d=4
theory associated to an $E_8$ singularity.

There are several ways to realize N=2 d=4 theories in string theory or M
theory. Let us consider cases
with a one-dimensional Coulomb branch for simplicity. One way is to wrap a type
II five-brane on a
curve $\S$. This picture can be related by a series of dualities to the type II
string on a Calabi-Yau
three-fold near a point of enhanced gauge symmetry. The theory at
low-energies on the five-brane is an N=2 d=4 theory \refs{\rsd, \rmoresd}. The
BPS states can be constructed in
terms of self-dual strings charged under the self-dual two-form $B$ wrapped on
a one-cycle of
$\S$. The d=4 BPS states then correspond to
geodesics on the Seiberg-Witten Riemann surface $\S$. A closely related
construction follows from
analyzing the strong coupling description of a system of type IIA four-branes
and five-branes
\rwittenM. At strong coupling, the system is better described in terms of an M
theory five-brane
wrapping a curve. The BPS states correspond to minimal area membranes with
boundaries on the
Riemann surface \refs{\randrei, \rpiljin}. These theories can also be
geometrically engineered \rgeom,
and the BPS states will correspond to branes wrapping various cycles of the
local geometry.

The approach that will primarily concern us in this paper is the probe
construction of the N=2 d=4
theory. Our probe is a D3-brane which is placed at a point on the moduli space
of the d=4 theory.
The singularities on the moduli space are replaced by seven-branes, some of
which are mutually
non-local. This picture can be derived from  F theory \rFtheory\ on $K3$ which
geometrically captures
the non-perturbative dynamics of an orientifold seven-plane \rsen. The
orientifold plane splits into
a collection of seven-branes, some of which we can send to infinity. In this
way, the resulting
theory is type IIB string compactified on $\CP{1}$ wih points deleted and with
a varying coupling
constant. The coupling constant is given precisely
by the solution of an N=2 d=4 theory. In the case studied by Sen, the d=4
theory was $SU(2)$ Yang-Mills
with four hypermultiplets. This result was given a very direct physical
interpretation by placing a
D3-brane probe at a point on the base space \rprobe. At low-energies, the
theory on the probe is $SU(2)$
Yang-Mills with four massive hypermultiplets, and the gauge coupling for this
d=4 theory coincides with
the type IIB string coupling. In a similar way, theories with no known
Lagrangian descriptions can be
constructed by placing the probe in the vicinity of more exotic singularities
such as an $E_8$ singularity
\rsunil.

Sen argued that certain BPS states in the probe theory can be realized by open
strings stretching from
the D3-brane to a seven-brane along a geodesic on the moduli space. This picture
was further explored in \rand. This
geometric realization of the BPS
states is very elegant and compelling physically, yet there was a mystery
associated to this picture. As
we shall explain, geodesics
involving single strings cannot realize all the BPS states of the probe theory.
Most of the states have
to be realized in a slightly more sophisticated way. The aim of this paper is
to explain how these states
are constructed, and to explore the relation between the various geometric
realizations of these BPS
 states in terms of membranes in M theory, open strings and self-dual strings.
Each of these pictures is
perhaps better suited for analyzing specific generalizations of this
construction. For example, the M
theory picture can be naturally generalized to the $SU(N)$ case, while the
probe approach can be
extended easily to the
case of $E$-singularities. Lastly, understanding BPS states in F theory
compactifications with D3-branes
has an immediate application to compactifications of F theory to four
dimensions, where one  of the
ways of
obtaining an anomaly free theory involves placing D3-branes in spacetime
\rfourfold.

While we were in the process of writing up our conclusions, some interesting
papers with  overlapping
results appeared \refs{\rima, \roren, \rzwie}. Some earlier work on
understanding the BPS spectrum
of the probe theory appeared in \ransar.

\newsec{Open String Realizations of BPS States}

\subsec{Single-string BPS states}

We wish to consider F theory on $K3$, which is equivalent to considering the
type IIB string with
24 seven-branes, some of which are mutually nonlocal. The positions of these
$(p,q)$ seven-branes in the
transverse two-dimensional space, parametrized by $u$, are encoded in the
choice of elliptically-fibered
$K3$. We can describe these spaces by specifying a curve,
\eqn\curve{ y^2 = x^3 + f(u) x + g(u),}
where $f$ and $g$ are polynomials in $u$ of degree $8$ and $12$, respectively.
The zeroes of the
discriminant of the curve,
\eqn\dis{\Delta = 4 f^3 + 27 g^2,}
describe the location of the seven-branes, while the monodromy of the complex
type IIB string
coupling around a particular zero encodes the $(p,q)$ charge of the seven-brane
located at that
point. When some of the seven-branes become coincident, the K3 can develop a
singularity. By placing
a D3-brane probe at a point on the base of the $K3$, we can induce a
non-trivial four-dimensional theory
on the probe \rprobe.

BPS states in the four-dimensional theory should correspond to string
configurations ending on the
D3-brane probe \rsen. Let us start by considering single string configurations.
Only $(p,q)$ strings
can end on a $(p,q)$ seven-brane; however, any string can end on the D3-brane.
Let us consider a
single D7-brane at a point $u_0$. On circling the seven-brane, the string
coupling
$$ \tau \sim {1\over 2\pi i}\ln(u-u_0), $$
undergoes a monodromy. As a result, we need to place branch cuts on the
$u$-plane to account for the
monodromy around each seven-brane. Let us place the D3-brane at a point $A$ on
the $u$-plane. A $(p,q)$
string stretching from the D3-brane along the $u$-plane on a curve $C$ has a
total mass given by,
\eqn\tension{\int_C | dw_{p,q} |,}
where $ dw_{p,q}$ is determined by the metric for the F theory background:
\eqn\defw{ |dw_{p,q}|^2 =  |p+q\tau|^2 | dw_{1,0}|^2. }
The holomorphic form $ dw_{p,q}$ is given by the integral of the 
holomorphic two-form,
$$ {dx du\over y}, $$
along the $(p,q)$ cycle of the elliptic fiber
in the locally flat basis in $H_{1}$ of the fibers. BPS configurations correspond 
to geodesics in the metric $ |dw_{p,q}|^2$ and
for these strings,
 $$ M_{p,q} = \int_C | dw_{p,q} | =  | \int_C dw_{p,q} |. $$
Let us choose a $(p,q)$ seven-brane at some point $B$. We can first inquire
about single-string
BPS states. These states correspond to geodesics starting at $B$ and extending
to the D3-brane at
$A$. How many such geodesics exist?

For a given geodesic $\g$, we can define a phase which is constant along the
geodesic
\eqn\phase{ \phi = \Arg \left[ w_{p,q}(A) \right] .}
Specifying $\phi$ determines the tangent vector to the curve $\g$ at $A$. This
uniquely determines the
geodesic which must terminate with finite length at the location of a $(p,q)$
seven-brane.
Therefore, all
single-string geodesics are uniquely fixed by the choice of central charge. Now
it is clear that we
have a problem. For example, the solution for the vacuum structure of $SU(2)$
N=2 Yang-Mills only
contains two singularities where a monopole or dyon becomes massless. In the F
theory picture,
these singularities are replaced by appropriate $(p,q)$ seven-branes. From the
argument
just presented, we only see BPS states that correspond to single strings
stretching from the
D3-brane to either of the two singularities, where the strings may pass through
a branch cut on the
$u$-plane. As we shall show in the following section, these strings only give a
finite number of
states. However, the semi-classical spectrum of pure $SU(2)$
Yang-Mills is easily determined. There are W-bosons with charges
$(\pm 2, 0)$ and dyons with charges $(2 n, \pm 1)$ where $n \in \IZ$. Higher
magnetic charge dyons
would correspond to holomorphic $L^2$ forms on monopole moduli space, but a
non-compact
Calabi-Yau manifold has no such forms \rmonopole.  How
are the remainder of these BPS states realized in string theory?

\subsec{Various routes to M theory}

To realize the full BPS spectrum, we need to consider multi-string
configurations \rschw. Classical string configurations do not generally correspond to
quantum mechanical BPS states. In the case
of maximal supersymmetry, the existence of such junctions as quantum mechanical
bound states was
argued in \rqthree\ using gauge dynamics and in \rmthree\ using the lift to M
theory. The dynamics of
these junctions has been explored further in \refs{\rsoojong, \rcallan}. In the
presence
of seven-branes, the question of whether a BPS junction exists is more subtle;
see \rmseven\ for a
discussion of non-terminating junctions in the presence of seven-branes. We
need to consider the
situation where legs of the junction terminate on seven-branes. In this case,
we need to determine a
selection rule for which junctions are BPS.

We need to consider configurations of string junctions with legs terminating on
three-branes
and seven-branes. To study when a string junction exists as a BPS
configuration, we will lift
the configuration to M theory. There are two ways of lifting F theory on $K3$
with a D3-brane
to M theory. For the first route, we consider F theory on $K3 \times S^1$ with
the $S^1$ transverse
to the D3-brane. Let us recall that M theory on $T^2$ is dual to the type IIB
string on a circle.
If the \kh class of the torus is $A$ then the radius $R$ of the circle
is given by,
$$ R = {1\over M_{pl}^3 A}, $$
where $M_{pl}$ is the eleven-dimensional Planck scale \rmIIB.
In this lift, the D3-brane is realized
in M theory as an M5-brane wrapping the torus $T^2$. In this way,
we connect with the pictures
presented in \refs{\rsd, \rwittenM} of a five-brane wrapped on a
curve $\S$. We are interested in the limit where $R$ is very large
and so the \kh class of the torus is very small. This limit is
essentially the opposite of the limit studied in \rwittenM, where the
\kh class was taken to infinity.

A $(p,q)$ stretched on
the $u$-plane and terminating on the D3-brane lifts to a membrane
wrapping the $(p,q)$ cycle of the
torus and stretched along the 1-cycle on the $u$-plane.
We will explore the relation between
the boundary of the membrane, which is composed of 1-cycles on $\S$, and
self-dual strings in more detail in the following section. Our starting
point is then a membrane wrapped on a curve in the total space of the elliptic
fibration. The boundary of
the membrane must lie on the curve $\S$. $\S$ is the elliptic fiber over the
point $A$
in the $u$-plane where our D3-brane was originally placed. The fiber is
holomorphic in a distinguished
complex structure, so the curve $\S$ is holomorphic in this particular complex
structure. If the
configuration is BPS then the world-volume of the membrane must be a
holomorphic surface with a boundary
on $\S$. For the world-volume to be a minimal surface, we need to adjust the
boundary so that the
world-volume intersects $\S$ at right-angles \refs{\randrei, \rpiljin}. If they
do not intersect
at right-angles then
although the membrane itself is a BPS configuration, the combined M5/M2 system
is not BPS \rfay. It
will turn out that not all string junctions lift to BPS configurations in the
total space.

The second route involves taking the D3-brane to be wrapped on the circle
$S^1$. In this case, the
D3-brane is realized in M theory as an M2-brane transverse to the $T^2$. The
end of the membrane
representing the $(p,q)$ string now looks like a point-particle in the
world-volume of the M2-brane.
If we further compactify a circle $S^1$ transverse to this configuration, we
can reduce the
configuration to a D2-brane intersecting a second D2-brane at a point. 
The intersection
point common to these two D-branes can be deformed to a more general curve, so
both branes should be considered part of a single brane. Since the 
D2-brane taken as the probe is significantly deformed by the intersecting
brane, it is no longer clear that there is a reasonable gauge theory 
interpretation for this configuration. 

This second approach corresponds to compactifying the four-dimensional theory
to three-dimensions on
a circle in the $x_3$ direction with radius $R$. The moduli space of the
uncompactified
four-dimensional theory is two-dimensional. On
compactification, we obtain two new compact scalars giving a four-dimensional
moduli space with
hyper\kh metric. One scalar $\phi_3$ corresponds to the choice of Wilson line
on the compact
circle while
the other direction is the expectation value of the scalar $\phi_D$ dual to the
resulting
three-dimensional gauge theory. In the limit where $R \r 0$, the circle
corresponding to $\phi_3$
decompactifies, and the moduli space is $ \IR^3 \times S^1$ equipped with a
hyper\kh metric.

\subsec{The relation with self-dual strings}

Another picture of BPS states in the N=2 theory  
follows from realizing the
 theory on an M theory or type IIA five-brane wrapping a  
Riemann surface $\S$ in a
six-dimensional space $\CC$ of $SU(2)$ holonomy.
The five-brane worldvolume theory has
$(0,2)$ superconformal invariance in flat spacetime. The  
matter
content of this theory comprises a tensor multiplet,  
which contains
five scalars describing the transverse coordinates of the  
brane
in eleven dimensions. The theory on the five-brane  
worldvolume
is twisted in the sense that the scalars become sections
of the normal bundle $N$ to $\Sigma$ in $\CC$. The  
condition for
$\CC$ to have $SU(2)$ holonomy implies that two out of  
the five scalars
become one-forms on $\Sigma$, while the rest are the  
ordinary functions.
Upon reduction to four dimensions the twisted scalars
give rise to $2g$ scalars, where $g$ is the genus of the  
Riemann
surface $\Sigma$. After reduction on $\S$, the tensor  
field $B^{+}$ gives $2g$
gauge fields
which include pairs of electric-magnetic duals. More  
precisely,
if $A^{i}, B_{i}$ is a basis for $H_{1}(\Sigma)$ then the  
two gauge-fields,
\eqn\ggflds{A_{i} = \int_{A^{i}} B^{+}, \quad {\tilde A}^{i} =
\int_{B_{i}} B^{+},}
are electric-magnetic duals. The three extra scalars
which are sections of the trivial part of the normal
bundle together with the integral of the two-form $B^{+}$  
over $\Sigma$
form a neutral hypermultiplet, which decouples from the  
rest of the fields.
In the D3-D7 picture, it describes the relative motion of
the D3-brane in the directions within the D7-brane  
worldvolume.  Moreover the
periodicity of the scalar coming from the $B^{+}$ field  
has to do with the fact that
in order to map the D3-D7 picture to M theory, we had to  
compactify one of the
directions along the D7-brane on a circle.

The Riemann surface $\Sigma$
has a meromorphic one-differential $\lambda$,
which is induced from the ambient space hyper\kh  
structure:
$\lambda = d^{-1} \omega_{c} \vert_{\Sigma}$.
Here $\omega_{c}$ is a holomorphic
symplectic form on $\CC$.
The five-brane theory contains self-dual strings which are
the boundaries of  M theory membranes. By wrapping such a
string along a cycle $\sigma \in H_{1}(\Sigma)$, we get a  
particle
$P_{\sigma}$
in the effective four-dimensional theory. The mass of the  
particle
$P_{\sigma}$
is given by the area of the membrane which has a curve in  
the homology
class $\sigma$ as a boundary. Arguments similar to  
\randrei\ show
that such a minimal membrane has
to be holomorphic in one of the
complex structures of $\CC$. The area of a holomorphic  
curve can be
computed as an integral of the \kh  form. The  
hyper\kh  manifold
$\CC$ has a two-sphere worth of complex structures.
In the complex structure $u \in \IC \cup \{ \infty \}$,  
the $(1,1)$ \kh form can be written as:
\eqn\kfrm{\omega_{r} (u) = {{1 - \vert u  
\vert^{2}}\over{1 + \vert u
\vert^{2}}} \omega_{r} + {{ i \bar u }\over{1 + \vert u
\vert^{2}}} \omega_{c} + {{-i u }\over{1 + \vert u
\vert^{2}}} \omega_{\bar c},
}
$\omega_{\bar c} \equiv \bar\omega_{c}$. 
Correspondingly, the $(2,0)$ form in this complex structure
is given by:
\eqn\hlfrm{\omega_{c}(u) = {1\over{1 + \vert u  
\vert^{2}}} \omega_{c}
- {{2\bar u}\over{1 + \vert u \vert^{2}}} \omega_{r} +  
{{\bar u^{2}}\over{1
+ \vert u \vert^{2}}} \omega_{\bar c}.}
Now given a two-surface $\Sigma_{\sigma}$ which is  
holomorphic in the
complex structure $u$, we may write:
$$
{\rm Area} \, \Sigma_{\sigma} = \int_{\Sigma_{\sigma}}  
\omega_{r}(u) =
 {{1 - \vert u \vert^{2}}\over{1 + \vert u
\vert^{2}}} A_{r} + {{ i \bar u }\over{1 + \vert u
\vert^{2}}} A_{c} + {{- i u }\over{1 + \vert u
\vert^{2}}} A_{\bar c},
$$
where $A_{r, c, \bar c} = \int_{\Sigma_{\sigma}}  
\omega_{r, c, \bar c}$. Due to
holomorphicity of $\Sigma_{\sigma}$ we have:
$$
0 = \int_{\Sigma_{\sigma}} \omega_{c}(u) = {1\over{1 +  
\vert u \vert^{2}}} A_{c}
+ {{2i  u}\over{1 + \vert u \vert^{2}}} A_{r} +  
{{u^{2}}\over{1
+ \vert u \vert^{2}}} A_{\bar c},
$$
hence
$$
{\rm Area} \, \Sigma_{\sigma} = \sqrt{A_{r}^{2} + \vert  
A_{c} \vert^{2}}
$$
Now consider varying the surface $\Sigma_{\sigma}$ keeping the
homology class of the boundary fixed. We get
$$
\delta A_{c} = \int_{\delta \p \Sigma_{\sigma}} \omega_{c} = 0,
$$
since the boundary is bound to lie on the fiber which is  
holomorphic
in the original complex structure corresponding to $u=0$
and moreover Lagrangian with respect to $\omega_{c}$.  
Hence
we cannot change $A_{c}$ and in order to minimize the  
area, we must make $A_{r}$ as small
as possible. The absolute minima would correspond to  
$A_{r}= 0$, which means
that $\Sigma_{\sigma}$ is holomorphic in the complex  
structure $u$ with $\vert u \vert = 1$, 
\eqn\fcmp{u = i {{A_{c}}\over{\vert A_{c} \vert}}. }

We can compare this description to that of self-dual  
string theory where the BPS states
are represented by
the geodesics in the metric $ds^{2} = \vert \lambda \vert^{2}$
in the given homology class:
\eqn\bps{
M_{P_{\sigma}} \geq \vert \oint_{\sigma} \lambda \vert .}
The particle $P_{\sigma}$ is charged under the $U(1)$ gauge
symmetry corresponding to $\sigma$. This follows from the  
fact that
the self-dual string is charged under $B^{+}$.
Particles which satisfy the bound \bps\ for a given  
choice of charges
correspond to BPS states. Therefore the BPS counting problem,  
at least in the cases where the minimal surfaces have  
$A_{r} = 0$, is equivalent to that in the theory of  
self-dual strings.
Note that the solutions of \bps\ can have
moduli. In this case, the moduli space needs to be  
described and quantum
mechanics on this space will determine the degeneracy of  
BPS states.

The relation between the picture of BPS states as geodesics
on the $u$-plane and that of geodesics on the Riemann
surface $\Sigma_{u}$ becomes clear once we lift
both pictures to M theory. Then we are really studying
minimal holomorphic surfaces representing the  
configurations of membranes.
In the limit where the fiber  \kh class is small, we have  
a good projection of the
membrane onto the $u$-plane giving an open string  
geodesic. By taking the boundary
of the membrane on $\S$, we obtain self-dual strings.

\subsec{A selection rule}

To derive the selection rule, we need to consider a three-string junction in
the neighborhood of a
D7-brane. The configuration is displayed in figure 2.1af. We need to know what
choices of $r$
and $(n,m)$ give BPS string configurations, where $n$ and $m$ are relatively
prime. To determine the
selection rules, we perform the lift
to M theory as described above with no D3-brane. We will keep $R$ large so the size 
of the M theory torus is small
and we have a well-defined projection onto the $u$-plane. If there were no
$(n,m)$ string as in figure
2.1b, then the state certainly exists since a fundamental string can end on a
D7-brane. This
configuration alone lifts to a holomorphic curve in the total space. To check
if the configuration
with the $(n,m)$ string lifts to a holomorphic curve, we can compute the
intersection number of
the curve for the fundamental string with the curve for the junction. If the
intersection number
is negative then the curve for the junction cannot be holomorphic because the
intersection number
of two holomorphic surfaces is non-negative.\foot{The self-intersection number
of a single curve
can be negative even if the curve is holomorphic. For example, a holomorphic
two-sphere in
$K3$ has self-intersection number -2.}

\kartinka{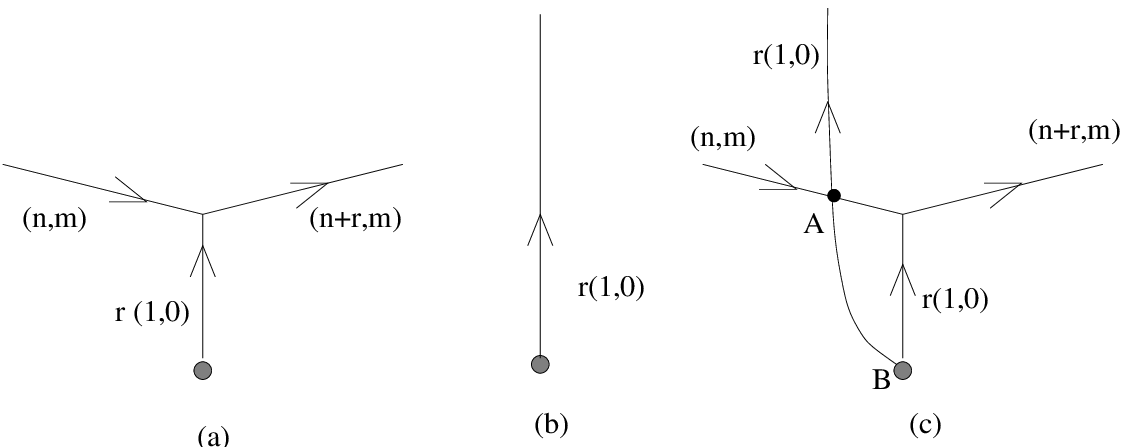}{Two cycles intersecting.}{2.1}

To compute the intersection number, we deform the cycle corresponding to the
fundamental string as
in figure 2.1c. This deformation does not change the intersection number. There
are two contributions
$i_A$ and $i_B$ to the intersection number from points $A$ and $B$. In general,
the intersection
number $i$ of an $(n,m)$ string with a $(q,p)$ string lifted to M theory is
given by the product of
the intersection number of the strings with the intersection number of the
corresponding cycles in
the torus. For the case shown in figure 2.2:

\risunok{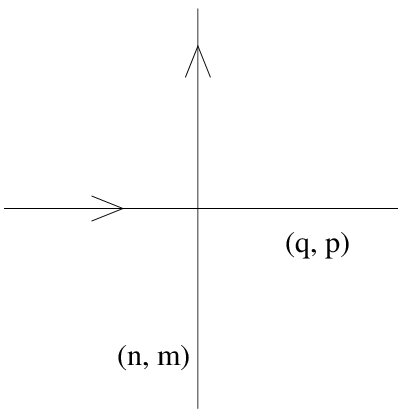}{Two strings intersecting.}{2.2}

\eqn\simpleint{i = p n - q m.}
{}From point $A$, we therefore obtain:
\eqn\ib{ i_A = r m. }
{}From point $B$, we obtain half the self-intersection number
of the cycle for the fundamental string,
\eqn\ia{ i_B = -r^2,}
and we require:
\eqn\equality{ -r^2 + m r \geq 0.}
It follows from \fcmp\ that the junction and the fundamental
string lift to membranes which are holomorphic in the
same complex structure; hence the argument of positivity
of the intersection index applies. This gives a selection rule for 
three-string junctions with a
leg terminating on a seven-brane.

There is a second way to see this selection rule. If the string
junction is deformed as drawn in
figure 2.3 using the monodromy,
$$  \pmatrix{ 1 & 1 \cr 0 & 1 \cr},$$
around a D7-brane then the configuration is no longer BPS
unless \equality\ is satisfied. To see this,
note that the orientation of the fundamental string in the transformed
configuration should be the
same as in the original configuration which implies that the
sign of $m+r$ must be opposite to the
sign of $r$. This gives the selection rule \equality.

\kartinka{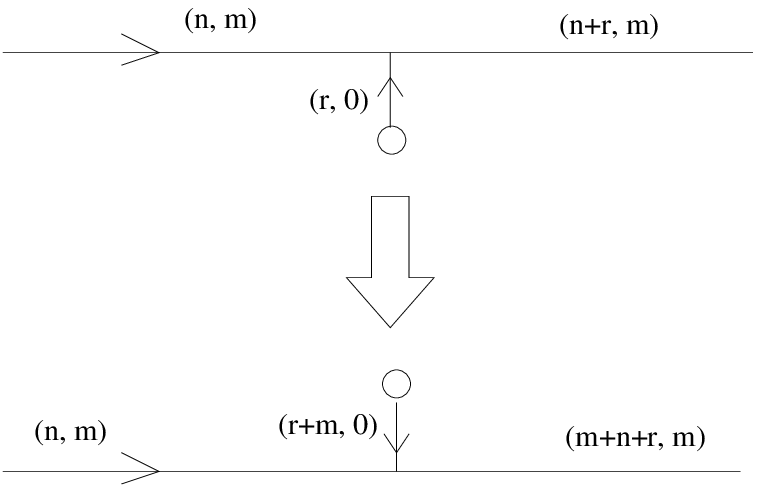}{Selection rule from deformation argument.}{2.3}

\subsec{Multi-string junctions and moduli}

In the case of string junctions with maximal supersymmetry where
there is only a flat metric on the
$u$-plane, there are actually moduli for the
junction. The simplest configuration is drawn in figure 2.4,
where the size of the triangle
represents the modulus. When the size of the triangle
goes to zero, we recover our usual
three-string junction. These moduli would appear, for example,
in the string configurations
discussed in \refs{\rorenym, \rymsolution}\ that describe $1/4$ BPS states 
in N=4 Yang-Mills.
In these cases, quantum
mechanics on the moduli space will give the multiplicity of BPS states,
and so the structure of the moduli space needs to be determined.

\kartinka{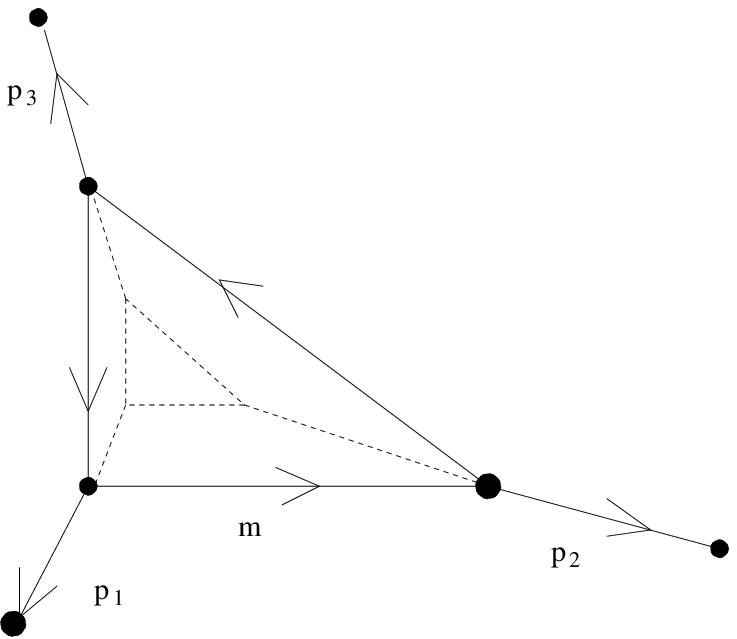}{A simple modulus for a single junction.}{2.4}

In general, the moduli space of a given junction consists of all string
networks
that have the same external legs. This can include gluing in a large,
but finite number of triangles.
Let us restrict to
the subspace of the moduli space where we blow up the vertex of the
three-string junction into
a single triangle. This is the case displayed in figure 2.4. The external
charges are fixed and
obey the usual constraint:
$$ \sum{ \vec{p}_i =0}, $$
where we label a $(p,q)$ string by the vector $\vec{p}$.
There are a discrete number of choices
for $ \vec{m}$ and these choices parametrize the components
of the moduli space. The allowed values of $ \vec{m}$
are restricted in that $\vec{m}, \vec{m} -\vec{p}_2$ and
$ \vec{m}+\vec{p}_1$ must form a closed triangle.
This amounts to the statement
that these vectors satisfy a linear relation with positive coefficients.
This condition is
geometrically realized in figure 2.5 by the constraint that the vector
$\vec{m}$ should lie
inside the triangle formed by the external vectors $\vec{p}$:
\risunok{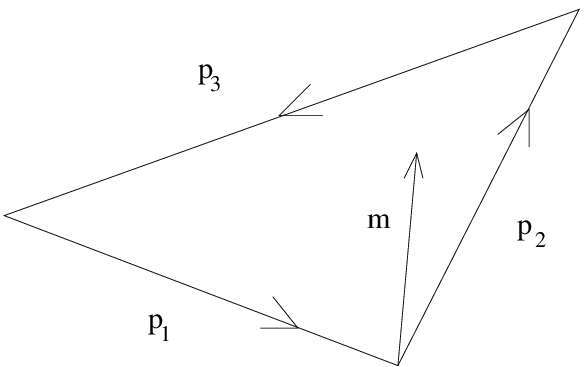}{Condition on the charge of
intermediate leg.}{2.5}

This tells us that the dimension of the
complete moduli space, including all possible internal triangles,
is given by the number of
integral points inside the triangle constructed from the external
vectors. Of course, the
moduli for junctions with more than three legs are described
in an analogous fashion. Similar
observations have been made in the study of five-dimensional
theories \refs{ \rnik, \rfivebranes}.

\subsec{Higher genus curves}

We can derive and generalize these results by considering the lift of the
string configurations
to M theory. Consider the space $ \,\IC^{*} \times \IC^{*}$ with coordinates
$(w,v)$
and let us view	it as a torus fibration over the base $\CU = \IR^{2}$:
$$
(w,v) \mapsto \left( x = {\rm log} \vert w \vert ,
y = {\rm log} \vert v \vert \right).
$$
Consider the holomorphic curve $C$ described by the equation
\eqn\expn{
\sum_{(m,n) \in \Delta_{C}} t_{m,n} w^{m} v^{n} = 0,}
where $\Delta_{C} \subset \IZ^{2}$ is just a set of exponents
$(m,n)$ for which $t_{m,n} \neq 0$.
For a given point $p = (x,y) \in \CU$, the intersection of $C$ with
the fiber $T^{2}$ over $p$  generically consists of a finite set of points.

Let us study the asymptotic behavior of $C$ along the trajectories
of the $\IC^{*}$ action labelled by a pair of mutually prime
integers $(\nu , \lambda)$:
$$
(w,v) = (\mu^{\lambda} w_{0}, \mu^{\nu} v_{0}),\quad \mu \in \IC^{*}.
$$
In the limit $\mu \to 0$, only those exponents $(m,n)$ in \expn\
will be important for which $- \left( m \lambda + n \nu \right)$
is maximal. If the set
$$L_{\lambda, \nu} = \{ (m,n) \vert
 - \left( m \lambda + n \nu \right) {\rm maximal} \}$$
contains more then one point then $C$ has an asymptotic component
pointing  in the $(\lambda, \nu)$ direction in the $\CU$ plane.
In fact, we may write down the equation for this component quite
explicitly. Let $a = -  \left( m \lambda + n \nu \right)$ be that maximal
value. We have:  $m = m^{\prime} a, n = n^{\prime} a$, for
$(m ^{\prime}, n^{\prime})$ mutually prime. 
The components corresponding to $(\nu, \lambda)$ are  labelled
by the points $\xi = w^{\nu}/ v^{\lambda}$
in the weighted projective space $\IC\IP^{1}_{\nu, \lambda}$
which solve the equation:
\eqn\cmpe{\sum_{(m^{\prime}, n^{\prime})\in {1\over{a}} L_{\lambda, \nu} }
t_{m^{\prime} a, n^{\prime} a} \xi^{m^{\prime} a/{\nu}}
= 0}
which describes an array of membranes which are wound  around a
$(p = \lambda, q = \nu)$ cycle of the fiber two-torus. 
The equation \cmpe\ implies that $\xi$
may assume only a finite number of values. For fixed
$\xi$ one gets a cylinder
of precisely  described form. Its projection to the $\CU$ plane is a
line, pointing in the $(p,q)$ direction. All such lines will be
parallel to each other, but need not be coincident.

The asymptotic behavior  of the curve $C$ is not changed if we add or
remove a point $(m,n)$ in the {\it interior} of $\Delta_{C}$.
Therefore the total number of moduli is given by the
number of integral points inside the convex hull of $\Delta_{C}$.
This is pictorially illustrated in figure 2.6. Notice that junctions
correspnding to non-convex polygons do not exist. A related discussion 
appeared in \rggm. Also note that we
have been discussing the case of maximal supersymmetry. For the situations
that we are primarily studying, we have less supersymmetry and these moduli
can be lifted. In section three, we will see cases of junctions smoothly
connected to single string configurations which cannot therefore have
any moduli.

\risunok{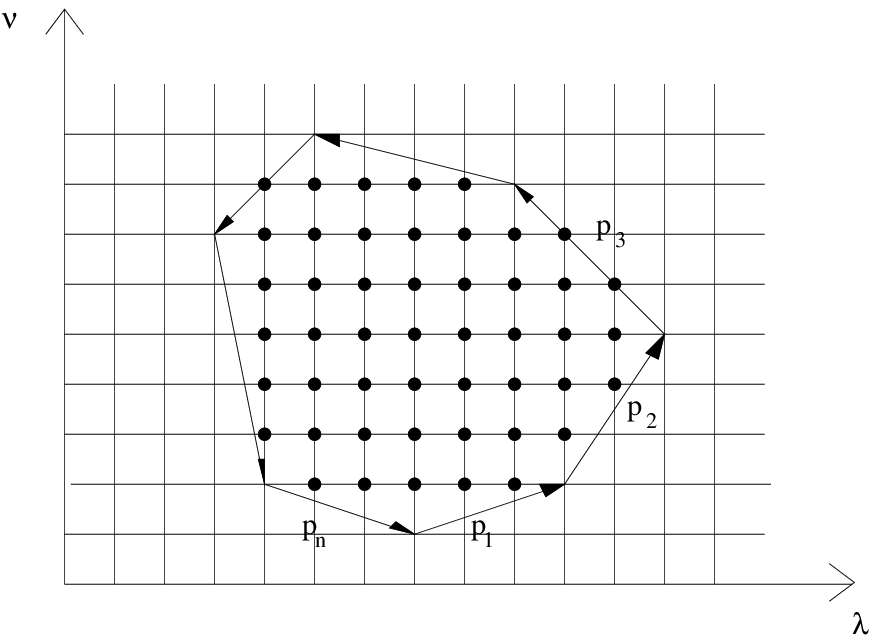}{Polyhedron associated to a multi-string junction.}{2.6}

\newsec{BPS states in $SU(2)$ Yang-Mills}

We will apply the preceeding discussion to the case of $SU(2)$ Yang-Mills
without matter. This is
the simplest situation
where non-trivial strong-coupling dynamics changes the BPS spectrum. The case
is easily constructed
as a probe theory by placing the probe in the vicinity of a deformed $D_4$
singularity and taking
the four mutually local D7-branes to infinity \refs{\rsen, \rprobe}. This
leaves two $(p,q)$
seven-branes with $(p,q)$ charges corresponding to
the two strong coupling singularities in the Seiberg-Witten moduli space. We
will normalize our charges
so that a W-boson has charge $\pm 2$. Then the charges for the two
singularities correspond to
$(\pm 2, 1)$ and $(0,1)$ depending on how we approach the singularity.

The moduli space has a curve of
 marginal stability (CMS) running
through both singularities shown in figure 3.1. Let us use the conventions for
the branch cuts
given in the second paper of reference \rbilal:
\risunok{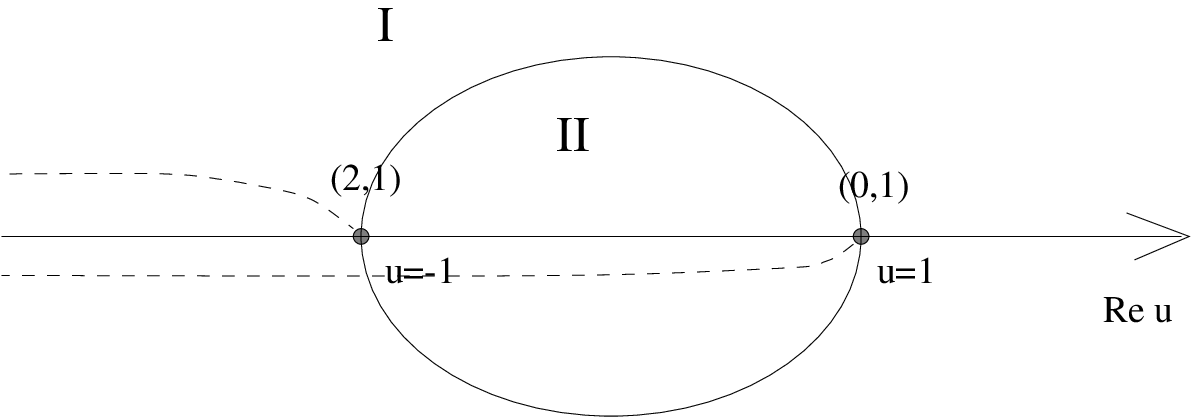}{The branch cuts.}{3.1}
The monodromies
around $u=-1,1,\infty$ are given by:
\eqn\monod{ M_{\infty} = \pmatrix{ -1 & 4 \cr 0 & -1 \cr}
\quad M_1 = \pmatrix{1 & 0 \cr -1 & 1 \cr}
\quad M'_{-1}  =\pmatrix{3 & 4 \cr -1 & -1}. }
In general, the monodromy around a $(p,q)$ seven-brane will be given by,
\eqn\pqmon{M_{p,q}=\pmatrix{ 1+pq & p^2\cr -q^2 & 1-pq \cr}.}
The only particles that should exist in the strong coupling regime, labelled
region II in figure 3.1,
are the particles which become light at the two singularities. That consistency
of the
proposed vacuum structure
requires all other states to decay when crossing the CMS is easily seen from
the monodromies in \monod.
Otherwise, by taking a semi-classical dyon around one of the two singularities,
we could generate a BPS
state which does not exist in the semi-classical region \rSW.

If we place our D3-brane probe at any point $u_0$ on the $u$-plane then we can construct
two canonical geodesics going from the probe to either the monopole or dyon singularity. 
These are simply straight lines in the $ w_{p,q}$ plane as we discussed in the previous
section. In addition, there can be another single string geodesic which crosses the
branch cut shown in figure 3.1 going from the dyon point to infinity. On crossing, the
charge of the string is acted on by the appropriate monodromy matrix. If the resulting
string has the same charge as either singularity, it can continue on and terminate at the
appropriate singularity. Recall that the initial trajectory of a $(p,q)$ string from the 
point $u_0$ is fixed by the charge of string. A case where there are two single string 
geodesics is shown in figure 3.2; the existence of this second geodesic can be verified 
either by numerical integration or analytically.   
\risunok{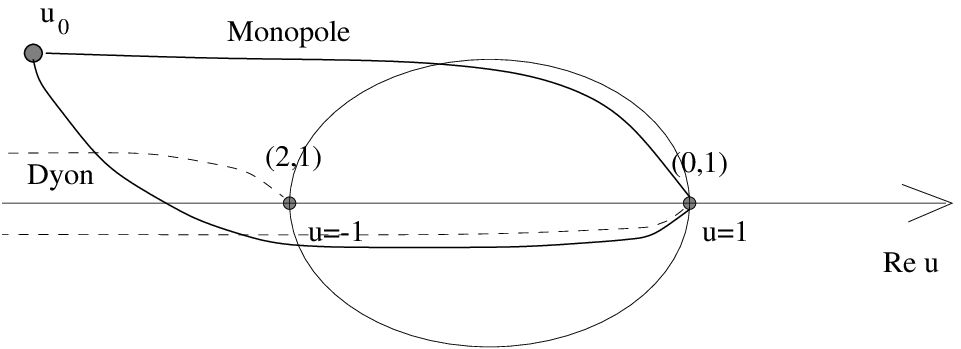}{Two single string BPS states.}{3.2}

If we move $u_0$ in figure 3.2 to the right, we need to drag the string ending on the
$(0,1)$ seven-brane through the other seven-brane. The D3-brane is far from the CMS
so the BPS state cannot decay. Rather it splits into a three-string junction drawn
in figure 3.3.

\risunok{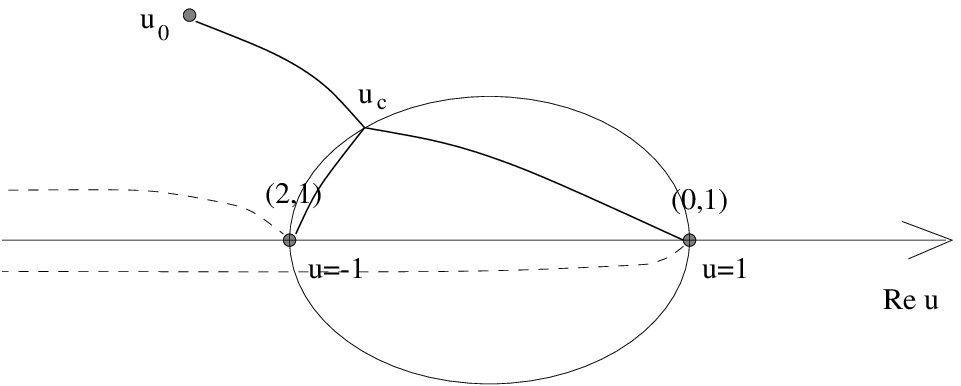}{A BPS three-string junction.}{3.3}
Under what conditions on $u_c$ is the junction BPS? Let us consider a junction with
charge $ \vec{p}_1$ on one leg and charge $ \vec{p}_2$ on the other leg so that the
final leg has charge $ \vec{p}= \vec{p}_1 + \vec{p_2}$. Then the mass of the state is given
by:
\eqn\mass{ \eqalign{ m(u_0) & = | w_{ \vec{p} }(u_0) | \cr
 &  = | w_{ \vec{p} }(u_0) - w_{ \vec{p} }(u_c) + w_{ \vec{p}_1 }(u_c) + 
w_{ \vec{p}_2 }(u_c)| \cr
 &  \leq | w_{ \vec{p} }(u_0) - w_{ \vec{p} }(u_c)| + |w_{ \vec{p}_1 }(u_c)| + 
|w_{ \vec{p}_2 }(u_c)|. }}
To satisfy the inequality, we require that
\eqn\phases{ \Arg \left[ w_{ \vec{p} }(u_0) - w_{ \vec{p} }(u_c) \right] =
\Arg \left[ w_{ \vec{p}_1 }(u_c) \right]  = \Arg \left[ w_{ \vec{p}_2 }(u_c) \right],}
which forces the point $u_c$ to lie on the CMS. This is rather beautiful and completely
in accord with our expectations. If we
bring the four local D7-branes that we sent to infinity back to the origin, the metric
on the moduli space becomes flat and the 6 seven-branes merge back into an orientifold
plane. As we expect, in this limit the string junction reduces to a $(p,q)$ string
stretching from the D3-brane to the orientifold plane. If we place our probe inside 
the CMS, we can have no BPS string configurations except single string configurations. 
The only geodesics correspond to the two canonical ones and we recover the expected 
non-analytic behavior of the BPS spectrum. 

So far, we have discussed the string junctions that arise by taking a single string 
configuration around infinity some number of times. This gives us a realization of
all dyon states with charge $(2n, \pm 1)$ where $n \in \IZ$ in terms of string
configurations. At first sight, it might seem that
these junctions have moduli but since they are continuously connected to a single string
configuration, it seems unlikely that any moduli actually exist. All these junctions 
necessarily exist but we can construct more junctions that satisfy 
\mass, but do not correspond to BPS
states in pure $SU(2)$ Yang-Mills. The condition of satisfying \mass\ is clearly
necessary but not sufficient.

To show that these extra states are actually not BPS, we
will apply a generalization of the selection rule from section 2.2. 
Note that we have all  dyons with
magnetic charge $\pm 1$. Suppose that we have
some dyon with higher magnetic charge. We denote
the central charge for this dyon by 
$$ x(2,1)+y(0,-1), $$
where $x$ and $y$ have to be of the same
sign for the state to be BPS. Without any loss of
generality, we may suppose that $x$ and $y$ are
both positive and coprime. 
We can then compute the intersection number of the curve
for this hypothetical dyon with the curve for
some known dyon with central charge
$$ s(2,1)+t(0,-1) $$
where $t=s\pm 1$.

After deforming the configuration as shown in figure 3.4\foot{
 The case shown corresponds to
${x\over y}<{s\over t}$. In the opposite case we should 
deform the $s,t$ cycle in the opposite direction.},
we compute the intersection index
\eqn\inters{
i=-(s-t)(x-y)+|sy-tx|.}
The formal rules for computing the intersection index of two
configurations with the same phase for their respective central charges
are the following: if they have legs
ending on the same seven-brane, with labels $x$ and $s$,
then we get a $-xs$ contribution (this is always
negative). If two legs intersect, then the contribution
is computed as in \simpleint\ (and this is always positive).

Notice that this intersection index does not change when
we change the phase of the central charge, even if
we cross the position of the seven-brane and the direction
of some leg is changed. This is what we expect since in
M theory language, this should correspond to a
smooth deformation of membranes. For example, consider
moving the junction in figure 3.4 to the right until
it crosses the monopole singularity. The labels of the states
are changed according to the monodromy transformation.
Explicitly, we have
\eqn\crossmon{\eqalign{
x(2,1)+y(0,-1)\to x(2,1)+(y-2x)(0,-1)\cr
s(2,1)+t(0,-1)\to s(2,1)+(t-2s)(0,-1)}}
Also, after we cross $u=1$, the intersecting legs will be
not $(s,y)$ as before, but $(\tilde{x},\tilde{t})$. 
Our rules for the intersection number then give us:
$$-|sx|-|(t-2s)(y-2x)|+|2x(t-2s)|=-sx-ty+2sy$$
which coincides with what we would have obtained from \inters.
Thus, the intersection pairing may be formally viewed
as giving us an invariant of the monodromy transformations.

Suppose that $x=y+b$ where $b>1$ (the case $b<-1$ can be considered in
an analogous way). Then take $s=t+1$. We can then compute the intersection 
number:
$$
i=-b+|y-tb|, $$
and we can always choose $t=\left[{y\over b}\right]$ to get
$i<0$. This proves that given the cycle corresponding to the
dyon with higher magnetic charge, we can always find a holomorphic
cycle whose intersection with the given cycle is negative. This implies
that we cannot realize it as a holomorphic curve.\foot{If $b=1$,
we would have to take $t=y$ to get $i=-1$, and this would imply
that our two cycles are actually the same. That is, we are 
computing the self-intersection number which does not
have to be nonnegative. Thus our argument does not rule out $b=1$.}
At first, we might worry that this argument is faulty because the
requirement that the intersection number be non-negative is only true
if we demand that both curves are holomorphic in the same complex
structure. However, in these cases, the direction of two legs for both
junctions agree and this determines the complex structure in which
their respective membranes must be holomorphic. These complex structures 
coincide so we can apply the intersection rule.  

The above argument does not rule out the case $b=0$ because $i$ given
in \inters\ is always non-negative for $x=y$. This case corresponds to 
a string junction representing a vector boson. The string leaving the 
D3-brane has charge $(\pm 2,0)$ and so represents two fundamental 
strings which must coincide on the $u$-plane by the BPS condition. 
The lift of this junction has a single complex modulus and there is
a single constraint coming from the known projection onto the $u$-plane.
This leaves a single real modulus in the M theory picture which should
be related to the modulus found in \refs{\randrei, \rsd}.  

\kartinka{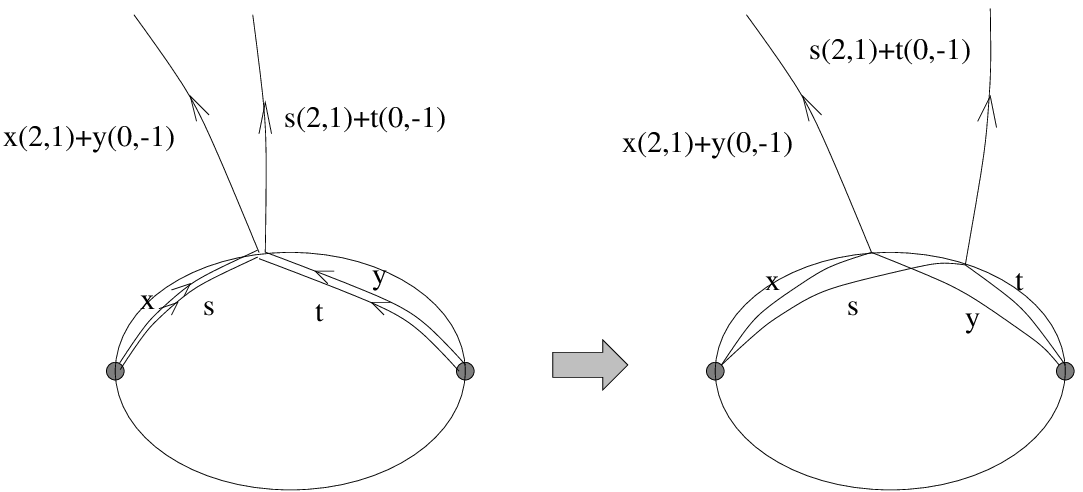}{Intersection of the cycle in question with
an existing cycle.}{3.4}

The existance of the modulus for the vector boson may also be seen
by computing the self-intersection number. Indeed,
for the configuration with charge $x(2,1)+y(0,-1)$, the
self-intersection number is  $-(x-y)^2$; this gives $0$ for
the vector multiplet ($x=y=1$) and $-1$ for monopoles and dyons 
($x=y\pm 1$). This suggests that hypermultiplets are represented
by disks, while the vector multiplet is represented by a cylinder.
A membrane with the topology of a cylinder has one real 
modulus, while the disk does not have any moduli. 
This picture is very different from what we would expect
for the infinite string junction in flat space. According to our 
description of moduli in section 2.5, we would
naively get approximately $\half n^2$ moduli for the dyon with
electric charge $n$. It turns out that a curved metric together with
the condition that the membrane end on a given fiber lifts all these
moduli.

This method of computing the BPS spectrum of N=2 theories should 
generalize to the case of multiple D3-brane probes which has been 
considered in \refs{\rmike, \rpyi}. It seems likely that the 
picture of geodesics on the moduli space of N=2 field theories is actually
more general and might well give a universal way of computing the BPS
spectrum, even for cases where no probe realization is known.

\bigbreak\bigskip\bigskip\centerline{{\bf Acknowledgements}}\nobreak

It is our pleasure to thank O. Bergman, A. Gorsky, 
 A. Hanany, A. Lawrence, F. Morgan, V. Sadov, A. Sen, C. Vafa
and E. Witten
for helpful discussions. N.N. and S.S. would also like to thank the Insitute
for Theoretical Physics in Santa Barbara
where part of this work was completed. The work of A.M. is supported in 
part by RFFI Grant No. 96-02-19086 and
partially by grant 96-15-96455 for scientific schools; that of N.N. by the
Harvard Society of Fellows,
 NSF grant PHY-92-18167, PHY-94-07194, RFFI grant No. 96-02-18046
and partially by grant 96-15-96455 for scientific schools; that of S.S. by NSF
grant DMS--9627351 and PHY-94-07194.

\listrefs

\end